\documentstyle[11pt,aas2pp4,psfig,flushrt]{article}
 
\begin{document}

\newcommand{\ivol}{\mbox{$\rm cm^{-3}$}}
\newcommand{\isup}{\mbox{$\rm cm^{-2}$}}
\newcommand{\isec}{\mbox{s$^{-1}$}}
\newcommand{\ten}[1]{\mbox{$10^{#1}$}}
\newcommand{\xten}[1]{\mbox{$\times 10^{#1}$}}
\newcommand{\wl}{\mbox{$\lambda$}}
\newcommand{\forb}[2]{\mbox{$[{\rm #1\, #2}]$}}
\newcommand{\ha}{\mbox{H$\alpha$}}
\newcommand{\hb}{\mbox{H$\beta$}}
\newcommand{\oiii}{\forb{O}{III}}
\newcommand{\oii}{\forb{O}{II}}
\newcommand{\oi}{\forb{O}{I}}
\newcommand{\nii}{\forb{N}{II}}
\newcommand{\sii}{\forb{S}{II}}
\newcommand{\kms}{\mbox{km~s$^{-1}$}}

\lefthead{Capetti et al.} 
\righthead{The origin of the NLR of Mrk 3}

\title{THE ORIGIN OF THE NARROW LINE REGION OF MRK 3: \\
AN OVERPRESSURED JET COCOON
\footnote{Based on
observations with the NASA/ESA Hubble Space Telescope, obtained at the
Space Telescope Science Institute, which is operated by AURA, Inc.,
under NASA contract NAS 5-26555 and by STScI grant GO-3594.01-91A}}

\author{A. Capetti}
\affil{Osservatorio Astronomico di Torino \\
Strada Osservatorio 20, 10025 Torino, Italy}
\author{D.J. Axon\altaffilmark{2}}
\affil{ Division  of Physical Sciences, University of Hertfordshire\\
College Lane, Hatfield, Herts AL10 9AB, U. K.}
\author{F.D. Macchetto\altaffilmark{2}}
\affil{Space Telescope Science Institute \\
       3700 San Martin Drive,       Baltimore, MD 21218}
\author{A. Marconi}
\affil{Osservatorio Astrofisico di Arcetri\\
Largo E. Fermi 5, 50125, Firenze, Italy}
\author{C. Winge\altaffilmark{3}}
\affil{Instituto de F\'{\i}sica\\
Universidade Federal do Rio Grande do Sul, Av. Bento Gon\c{c}alves\\
9500, C.P. 15051, CEP 91501-950, Porto Alegre, RS, Brazil.}
\altaffiltext{2}{Affiliated to the Astrophysics Division, Space Science
Department, ESA}
\altaffiltext{3}{CNPq Fellowship, Brazil}

\authoremail{capetti@to.astro.it,axon@stsci.edu,macchetto@stsci.edu,marconi@arcetri.astro.it,winge@if.ufrgs.br}

\begin{abstract}

We have obtained HST FOC f/48 long--slit optical spectroscopy of the
inner 2\arcsec\ of the Narrow Line Region of the Seyfert 2 galaxy Mrk 3
with a spatial resolution of 0\farcs06.  Spectra were taken in six
locations with the slit approximately perpendicular to the radio-axis.

In the region cospatial with the radio-jet, where the brightest
emission line knots are located, the velocity field is highly perturbed
and shows two velocity systems separated by as much as 1700 \kms. In
several locations the split lines form almost complete velocity
ellipsoids implying that we are seeing an expanding shell of gas. 
The diameter of this shell ($\sim$ 200 pc) is much larger
than the width of the radio-jet ($ d < 15$ pc). 
We interpret this to be the consequence of the rapid expansion of a cocoon of
hot gas, shocked and heated by the radio-emitting outflow, which 
compresses and accelerates the ambient gas. The cocoon 
mediates the energy exchange between jets and line emitting gas. 
The gas motions within
the NLR of Mrk 3 are therefore clearly dominated by the interaction between the
jets and the interstellar medium and the NLR itself is essentially a
cylindrical shell expanding supersonically.

With its current size of 200 pc the cocoon has expanded to several disk
scale heights. Due to the external gas density stratification, the hot
gas located above the plane of the disk blows out into the halo,
puncturing the bubble and fracturing the velocity ellipsoids. The
system is effectively momentum driven.

From the size and velocity of the expanding region, we derive
an upper limit to the radio-source age, $\lesssim 1.5 ~10^5$ years, and
a lower limit for the jet power, $\gtrsim 2~10^{42}$ erg s$^{-1}$,
required to inflate the cocoon and estimate that the jet minimum
advance speed is $3 ~10^{-3}$ pc per year. The total kinetic energy of
the high velocity gas associated with the radio-jet can be estimated as
$\sim 6~10^{54}$ erg, comparable to the total energy carried by the jet
over its lifetime and this quantitatively supports the idea that the NLR gas
is accelerated by the jet. 

Radio-outflows are associated with at least 50 \% of Seyferts galaxies
with typical sizes smaller than a few kpc. If the advance speed of Mrk
3 is representative of the Seyfert population then these sources must
also be short lived and probably recurrent.  Evidence that this is
indeed the case is provided by the fact that the expansion time-scale
derived for NGC 1068 is comparable to that seen in Mrk 3.

The jet kinetic luminosity of Mrk 3 is between 2 and 3 orders of
magnitude smaller than that derived for radio-loud AGNs with similar
emission-line luminosity. On the other hand, the fraction of jet power
dissipated in radio-emission is similar. We speculate that the main
distinction between radio-quiet and radio-loud AGN is ascribed to a
difference in jet power rather than to a different efficiency in
synchrotron emission production.

\end{abstract}


\section{Introduction}

During the last two decades ground based studies of Seyfert galaxies have firmly established that a close relationship exists between radio and line emission. In particular they showed that their Narrow Line Region (NLR) is 
inevitably cospatial with the radio-emission ( e.g. Wilson \& Ulvestad, 1983; Haniff, Wilson \& Ward, 1988). Furthermore, galaxies harboring a linear radio source have unusually wide line profiles (Whittle 1985) and display a correspondence between the location of the high velocity NLR gas with that of the radio ejecta (e.g. Whittle et al. 1988; Baldwin, Wilson \& Whittle 1987; Pedlar et al. 1989).
This spatial and kinematical association has been interpreted as due to 
the compression and acceleration of the interstellar gas induced by the expansion of the radio-ejecta (Wilson \& Willis 1980, Booler et al. 1982, Pedlar et al. 1989, Taylor et al. 1989, 1992). 

Hubble Space Telescope (HST) allowed us for the first time to fully resolve the region from which the narrow emission lines originate. The results of extensive narrow-band imaging of Seyfert galaxies clearly showed that the NLR morphology is completely determined by the presence of radio outflows (e.g. Bower et al. 1994, 1995, Capetti et al. 1995a, 1995b, 1996, 1997a, 1997b, Falcke et al. 1996, 1998). In particular, Seyfert galaxies with a lobe-like radio morphology have bow shock shaped emission line regions while those with a jet-like radio structure have jet-like emission line structures. These observations provide further compelling evidence for strong dynamical interactions between the NLR gas and radio-emitting ejecta. 

In this framework it is of great interest to study the effects of this interaction on the NLR dynamics at a high spatial resolution. This prompted a program of long-slit observations using the Faint Object Camera (FOC) in its spectrographic mode in order to study in detail the velocity field of the NLR of Seyfert galaxies. The first results from
observations of NGC 4151 (Winge et al. 1997, 1998) and NGC 1068 (Axon
et al. 1998) show clear evidence that the highly perturbed and complex
velocity field of the NLR of these objects is strongly influenced by
the jets propagation. In particular, high velocity gas components are
associated univocally with the jet region.

We now present the results of long-slit HST/FOC spectroscopy of Mrk 3, a Seyfert 2 galaxy which represents one of the clearest examples of close association between radio and line-emission. Radio-images show a linear structure extending
over $\sim 2$ \arcsec, dominated by  two symmetric and highly
collimated jets (Kukula et al. 1993). HST emission-line images of
Mrk 3 revealed that its NLR has a striking ``S''
shaped morphology composed of a series of knots, sheets or filaments
and is basically cospatial with the radio-jets (Capetti et al.
1995a, 1996). 

In this paper we show that the NLR of Mrk 3 is essentially an expanding shell surrounding the radio-jet whose velocity field is determined by the energy input of the outflowing jets. The detailed spatial and velocity information derived allow us to investigate the energetics of the jet/NLR interaction
and to set constraints on the properties and on the evolution of the
radio-source associated to Mrk 3.
We defer the discussion on line ratios and the ionization properties
of the emitting gas to a forthcoming paper.

Throughout this paper we will adopt a recession velocity for Mrk3 of
4000 \kms\ (z = 0.0133) which, for $ H_\circ = 50$ Km s$^{-1}$
Mpc$^{-1}$, yields a distance of 80 Mpc where 1\arcsec corresponds to
390 pc.

\section{Observations and Data Reduction} 
Mrk 3 was observed
using the FOC f/48 long--slit spectrograph on December 9$^{\rm th}$,
1996. One pixel corresponds to 1.78\AA\ and 0\farcs0287 along the
dispersion and slit directions, respectively.  The F305LP filter was
used to isolate the first order spectrum which covers the 3650--5470
\AA\ region and therefore includes the \oii\wl 3727, \hb\wl 4861 and
\oiii\wl\wl 4959, 5007 \AA\  emission lines.  An interactive
acquisition 1024x512 zoomed image was obtained with the f/48 camera
through the F342W filter to accurately locate the brightest
line-emission regions.  The slit, 0\farcs063x13\farcs5 in size, was
placed at a position angle of -25$^\circ$. Spectra with exposure times
of 1000 seconds were taken in the 1024x512 non--zoomed mode at 7
locations separated by 0\farcs4. An additional spectrum of 780 s was
taken immediately following the interactive acquisition image, before
the telescope small angle maneuver (position 2 in Fig. 1).

The data reduction follows the procedure described in detail by
\cite{m87}. In summary, all frames, including those used for subsequent
calibrations, were geometrically corrected by using the equally spaced
grid of reseaux marks etched onto the first photocatode in the
intensifier tube.  Any remaining internal distortion was corrected
tracing the spectra of two stars in the core of the globular cluster 47
Tuc.  The distortion along the spatial direction was obtained in a
similar way, tracing the brightness distribution of the emission lines
of the planetary nebula NGC 6543. Observations of NGC 6543 were also
used to obtain the wavelength calibration. The instrumental broadening
is estimated to be $\sim 320 \pm$ 20 \kms. Flux calibration was
performed using the observations of the spectrophotometric standard
star LDS749b.

To accurately determine the location of the slits we compared the
surface brightness profile derived from the HST FOC, f/96, \oiii\ image
of Capetti et al. (1996) with that measured from the spectra at the six slit
positions which yielded useful data.  The best match is displayed in
Fig. 1 and is accurate to within half a slit width
($\simeq0\farcs03$).  The slit locations, identified as POS1 through
POS6, cut across the emission-line region of MRK 3 in a direction
almost perpendicular to that of the radio-axis, P.A. +84$^\circ$
(Kukula et al. 1993).

\begin{figure*}
\centerline{\psfig{figure=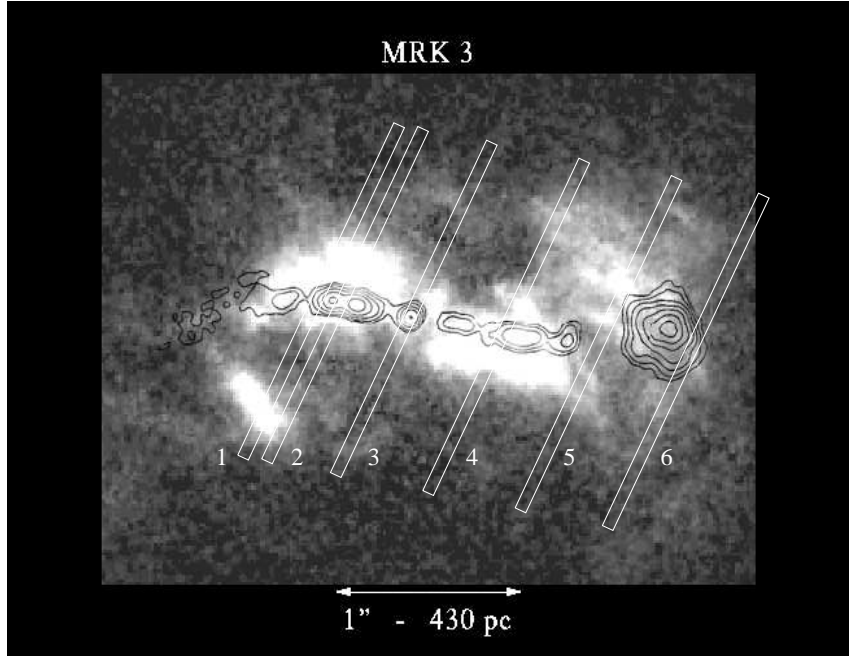,width=0.64\textwidth,angle=270}}
\figcaption{\label{fig:slitpos}HST/FOC image of Mrk 3 in the [O III] emission line from Capetti et al. (1996) with superposed the contour radio image from Kukula et al. (1993) and the 6 slit positions where the FOC f/48 spectra were taken. North is up and East is to the left.}
\end{figure*}

\section{\label{sec:results} Results}

Velocities were derived by fitting gaussians profiles to the
\oiii\ 5007$\lambda$ line at each individual pixel using the task
LONGSLIT in the TWODSPEC FIGARO package (Wilkins \& Axon 1992). In
several locations the line is composed of more than one component and
in these cases multiple gaussian fitting was performed. In regions of low
signal-to-noise, up to three pixels along the spatial direction were coadded.  The resulting velocity profiles across the slit at each location are shown in Figs. 2 through 8, where we also plot line widths and intensities.  The origin of the
X-axis for each slit location is set at the intersection with the radio-axis.

\subsection{\label{sec:spectra} Description of the individual spectra}

\noindent
$\bullet$ POS1:  the line profiles are split into two distinct velocity
systems in the region of the radio-jet, with a separation of $\sim$
1000 \kms. Toward the  South, only the redshifted component is visible.
The region of high velocity is extended over $\sim$ 0\farcs5.  The slit
also crosses the bright Southeastern blob. This blob is detached from
the main NLR structure and not directly associated with the radio-jet.
The gas in this region has a velocity of $\sim$ 3600 \kms\ and is
significantly blueshifted from the velocity on the East side of the NLR which is$\sim$ 4300 \kms\ (Wagner 1987, Metz 1998).
Furthermore the lines are quite broad, $\sim$ 700 \kms.

\noindent $\bullet$ POS2: this slit position was obtained during the
interactive acquisition and was located close to POS1. On the North
side of the NLR the velocity is approximately constant around 3900
\kms. Along the radio-axis there is a dramatic split of the lines with a
separation which increases very rapidly and reaches a maximum of
1700 \kms. Unlike in POS1, the velocity field forms nearly a
complete velocity ellipsoid. We are seeing an expanding
shell of size $\sim $ 0\farcs3 around the radio-jet. The line widths of
the individual components are always large, 500 \kms, and significantly
wider than the instrumental broadening of 320 \kms. The large increase
in line width just North of the split-line region is probably due to
spatial confusion of the two velocity systems at the periphery of the
shell.  On the opposite side of the jet the slit grazes the bright SE
blob. A smooth gradient of 200 \kms\ can be seen over a distance of
0\farcs4 and, again, line widths are large $\sim$ 700 \kms.

\noindent $\bullet$ POS3:  the slit is located on the brightest blob of
the NLR of Mrk 3, close to its center of symmetry and, likely, to its
hidden nucleus. The velocity field is remarkably flat on the blob  but
the lines are characteristically broad (500 \kms).  The emission from
this blob is redshifted by about $\sim$ 200 \kms\ with respect to the
galaxy systemic velocity.  Immediately North of this blob,
corresponding to where the radio-jet intersects the slit, lines are
again split in two components with a velocity separation of $\sim$ 1000
\kms.

\noindent $\bullet$ POS4: this location is symmetric to POS1 with
respect to the nucleus and the velocity field is very similar but  with
a reversed symmetry. Again it shows a very large line splitting (1200
\kms) along the radio-jet over $\sim$ 0\farcs4 in the form of a broken
velocity ellipsoid.

\noindent $\bullet$ POS5:  the gas kinematics are complex. Three highly
perturbed regions  correspond to three bright filaments crossed by the
slit. The first, toward the South, is part of the main NLR structure. The
remaining two, which are blueshifted by $\sim$ 600 \kms\ and have large
line widths, are associated with the radio-lobe. These structures are
superposed onto a more regular pattern determined by the diffuse
emission. Such complex line profiles has been seen previously on the
lobe of the powerful radio-Seyfert IRAS 04210+0400 (Holloway et
al. 1996).

\noindent $\bullet$ POS6: slit location 6 grazes the edges of the NLR
and of the Western radio-lobe in a region of diffuse line emission.
While the velocity field is essentially quiescent, with a dispersion of
less than 250 \kms, the line profiles are significantly broad only in
the region along the radio-axis, with a FWHM as high as 1000 \kms.

\begin{figure}
\centerline{\psfig{figure=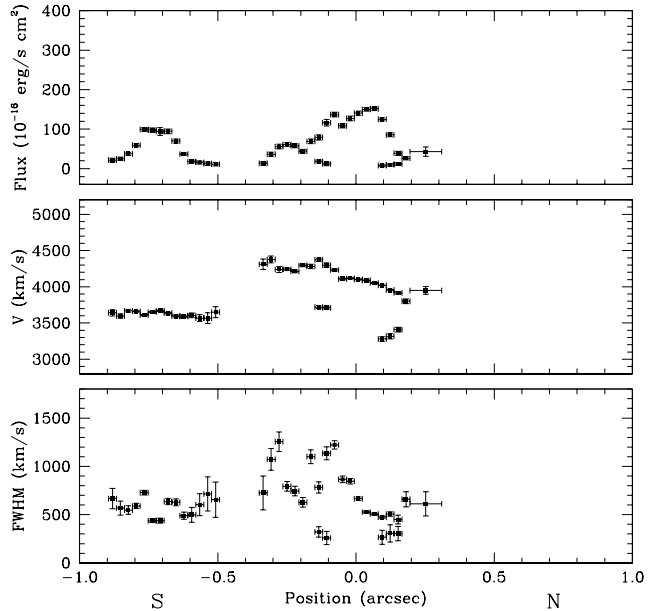,width=0.5\textwidth,angle=0}}
\figcaption{Intensity (upper panel), velocity (middle panel) and line widths (lower panel) measured at the slit position POS1.}
\end{figure}

\begin{figure}
\centerline{\psfig{figure=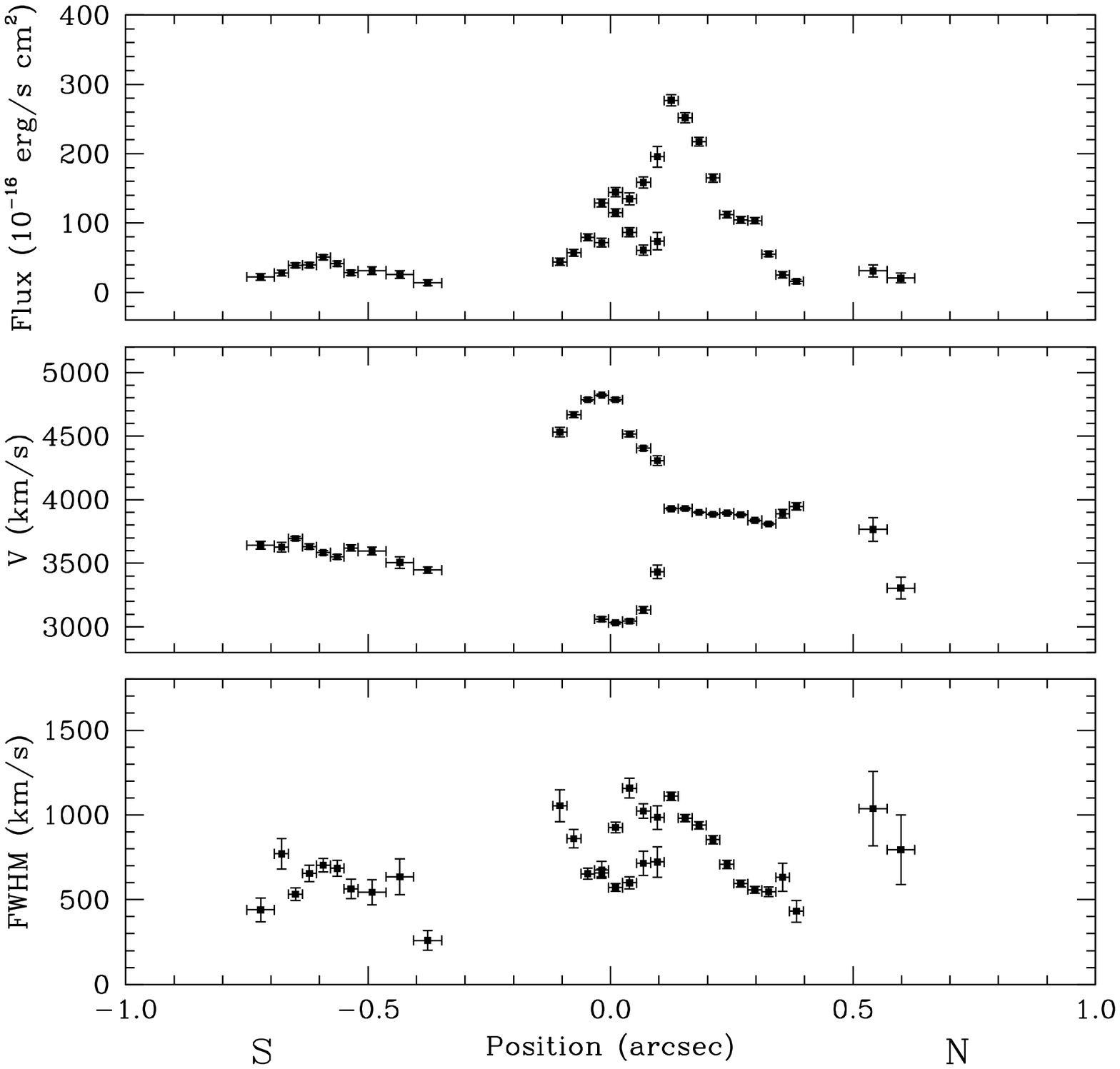,width=0.5\textwidth,angle=0}}
\figcaption{Same as Fig. 2 for slit position POS2.}
\end{figure}

\begin{figure}
\centerline{\psfig{figure=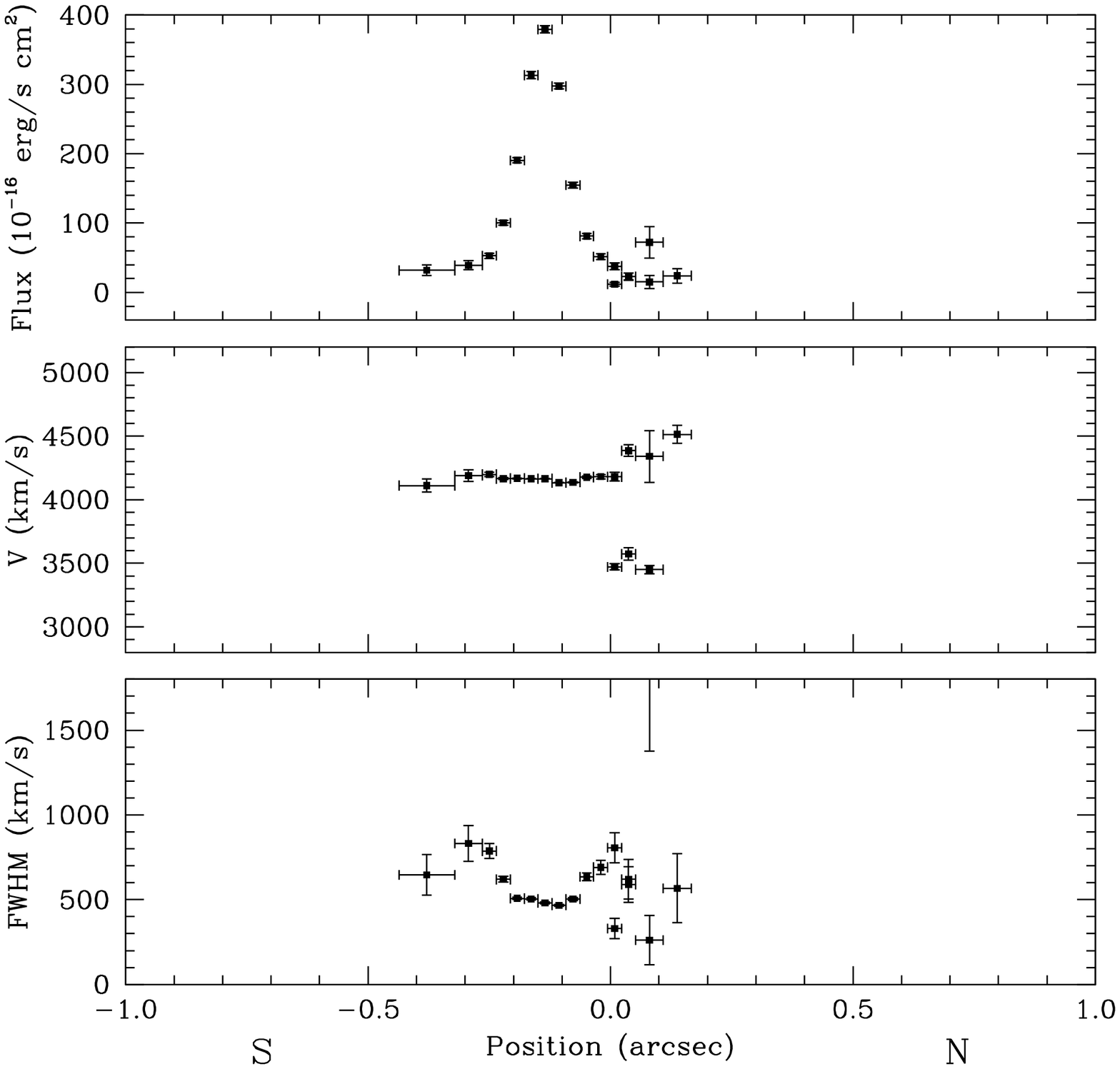,width=0.5\textwidth,angle=0}}
\figcaption{Same as Fig. 2 for slit position POS3.}
\end{figure}

\begin{figure}
\centerline{\psfig{figure=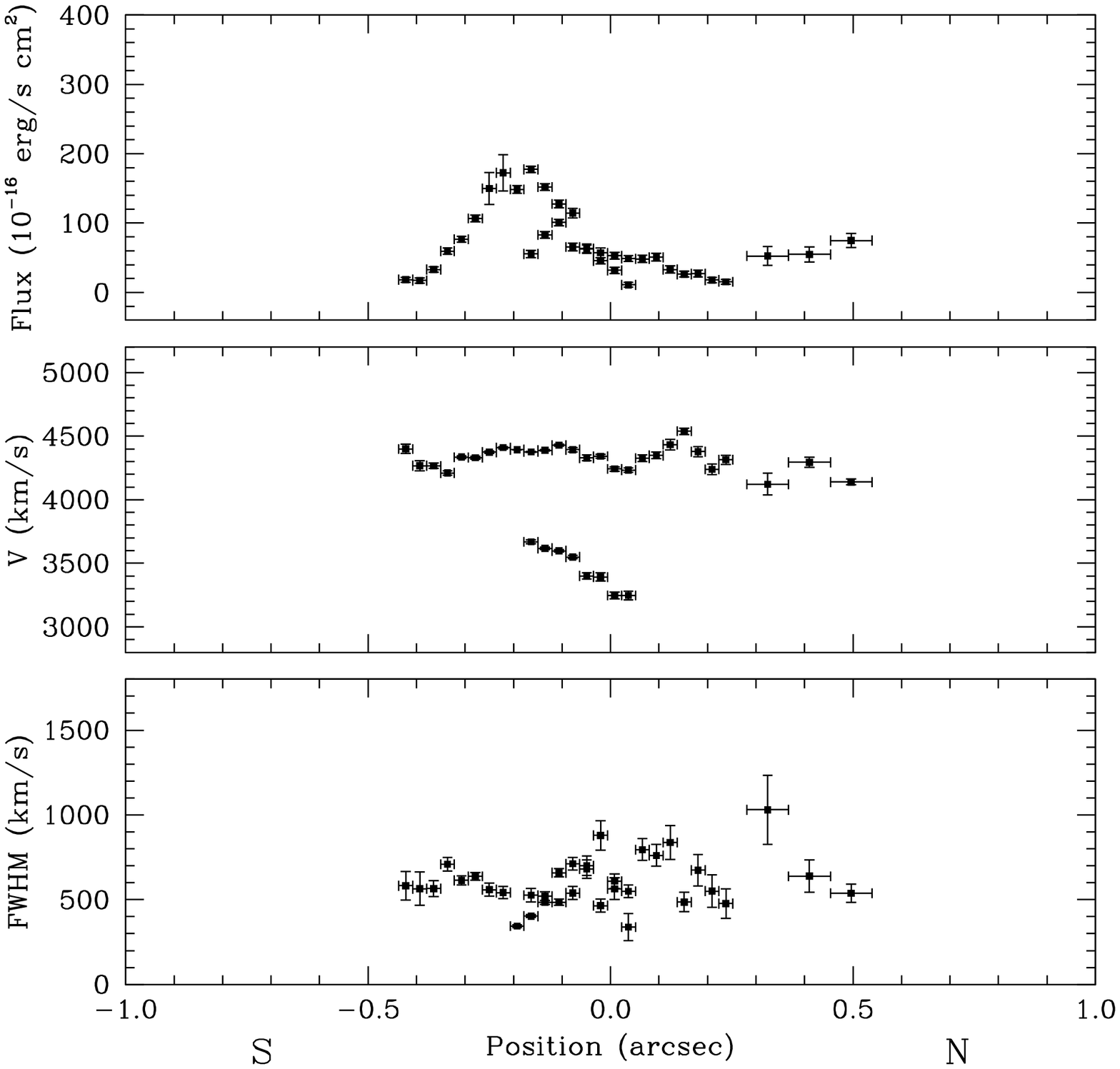,width=0.5\textwidth,angle=0}}
\figcaption{Same as Fig. 2 for slit position POS4.}
\end{figure}

\begin{figure}
\centerline{\psfig{figure=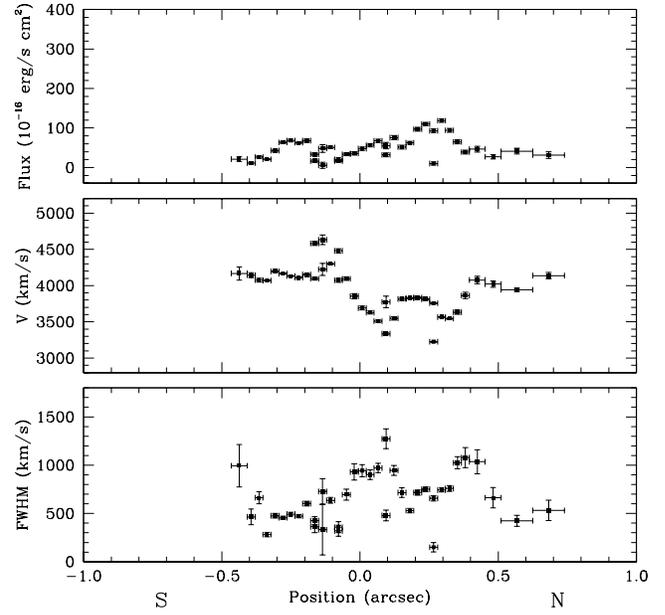,width=0.5\textwidth,angle=0}}
\figcaption{Same as Fig. 1 for slit position POS5.}
\end{figure}

\begin{figure}
\centerline{\psfig{figure=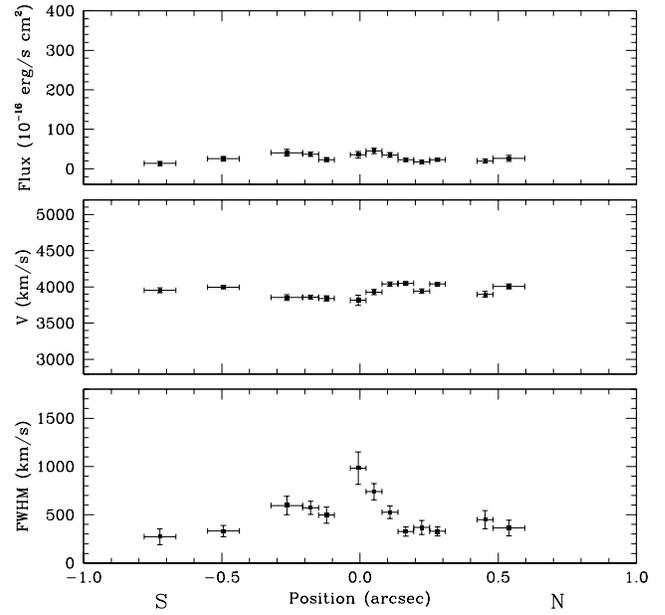,width=0.5\textwidth,angle=0}}
\figcaption{Same as Fig. 2 for slit position POS2.}
\end{figure}

\section{Kinematics of the NLR gas} 

The most dramatic result is that,
kinematically, the NLR of Mrk 3 takes the form of a broken velocity ellipsoid
centered on the radio-jet. This is the expected behaviour of an
expanding cocoon of gas circumscribing the radio-jet and is
unambiguous evidence that the NLR is created by the interaction of the
radio-jet through the interstellar medium.  Note that while there is a
misalignment between the NLR and radio axis (Capetti et al. 1995a) the
velocity ellipsoid is centered firmly on the jet.  Line widths in the
expanding regions are always broad, 500 \kms, suggesting significant
turbulence within the cocoon.

The diameter of the emission line shell ($\sim$ 200 pc) is much larger
than the radio-jet which is unresolved in the MERLIN images ($ d < 15$
pc). This is an indication that the radio-plasma does not interact {\sl
directly} with the line emitting gas. This is physically expected since
for a shock velocity of $\gtrsim$ 1000 \kms\ the post shock temperature
will be of $\gtrsim 10^7$ K (Taylor et al. 1992).  A hot high-pressure
cocoon then develops around the advancing radio-jet and mediates the energy
exchange between jets and line emitting gas.  At these temperatures the
main cooling mechanism is free-free radiation rather than line
emission. Strong support for this interpretation is the direct
detection of the ultraviolet counterpart to the Mrk 3 jet (Axon,
Capetti \& Macchetto 1998). 

The excess pressure of the hot gas
initially drives the supersonic NLR expansion.  But, as we shall now
explain, we believe that the NLR is currently in a momentum driven
phase as the bubble has been punctured.
The two key facts leading to this conclusion are the mirror symmetry
about the nucleus of the NLR emissivity and fractures in the velocity
ellipsoids.  As described above, on the Western jet the bubble is open
towards the South and the peak of emission is to the North (POS1 and
POS2) while the reverse is seen along the East jet (POS3 through
POS5).  This behaviour can be explained if the jets are propagating in
the stratified gas of a disk galaxy and the jets are inclined  with
respect to the gas disk (see Fig. 8). The dynamical evolution is very
similar to that of the wind driven superbubbles associated to OB
associations or supernovae in the ISM (Tomisaka \& Ikeuchi 1986, Mac
Low \& McCray 1988). Initially the bubble will expand asymmetrically
with respect to the jet axis with the side away from the disk expanding
more rapidly. It will still appear as a closed, although asymmetric,
velocity ellipsoid. The bright emission line region corresponds to the
side of the bubble which is snow-plowing into the disk.  At this stage
the NLR motions are pressure driven.  When the bubble grows to be large
enough it is more profoundly influenced by the stratification of the
galactic disk. With the current size of 200 pc the cocoon has probably
expanded to several disk scale heights and the hot gas located above
the plane of the disk is blowing out into the halo, puncturing the
bubble.  At this time the system is effectively momentum driven.

\begin{figure*}
\centerline{\psfig{figure=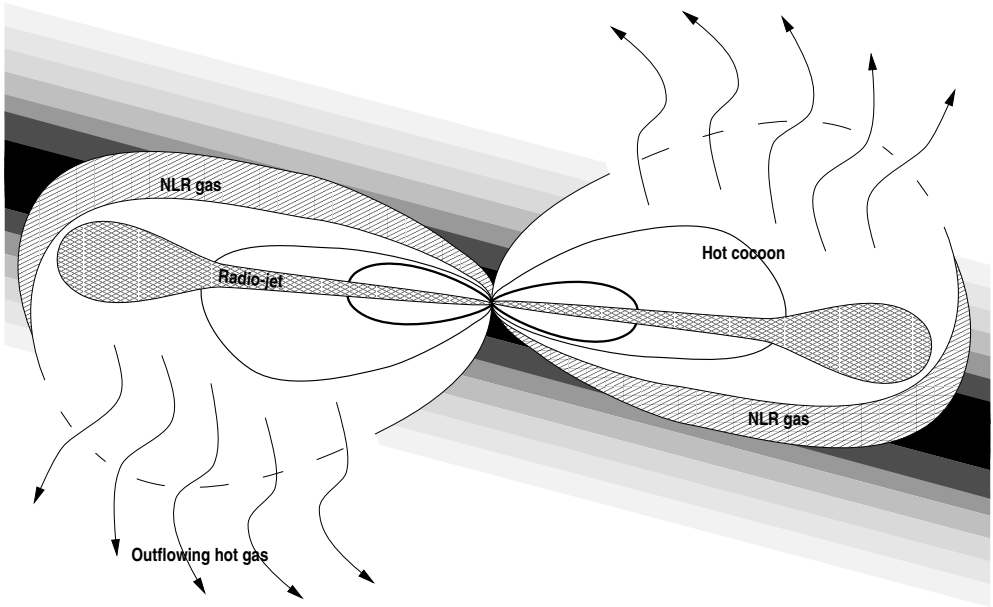,width=0.75\textwidth,angle=270}}
\figcaption{
Cartoon illustrating the effects of the relative jet/disk orientation on the cocoon expansion and on the NLR morphology.}
\end{figure*}

Aside from the highly perturbed gas on which we have concentrated so
far, at all slit locations there are extended regions in which the gas
is relatively quiescent, i.e. with a velocity field characterized by
small velocity gradients, small offset from the galaxy systemic
velocity and relatively narrow line profiles. This gas is probably
associated with the galaxy disk. The presence of an extended rotating
gas disk in Mrk 3 is clearly seen in ground based data (Metz 1998) and
it is normally observed in Seyfert galaxies. A similar two-components gas structure in the NLR has been recently seen also in NGC 4151 (Winge et al. 1998). Radio outflows disrupt only locally the ordered rotation pattern.

\medskip \section{\label{sec:cocoon} Evolution of the cocoon} 
The evolution of an overpressured cocoon driven by a supersonic jet has
been studied by several authors in the framework of powerful
radio-galaxies.  The same approach can be applied to the expanding
cocoon in Mrk 3 but, in this case, we can directly measure its volume
and expansion velocity which turn out to be crucial parameters to
constrain the model.

Begelman \& Cioffi (1989) showed that the evolution of a radio-source
is determined by the balance between the ram pressure of the external
medium with the thrust of the jet and, where the lateral expansion is
concerned, with the cocoon internal pressure, i.e.  $p_c \sim \rho_a
v_{c\circ}^2$, where $p_c$ and $v_{c\circ}$ are the cocoon pressure and
the present expansion velocity respectively and $\rho_a$ is the
external density. The medium in which the jet propagates is likely
to be clumpy; a radio-source evolves within a non-uniform medium in a
way similar to that with which it would interact with a smooth medium with the
same average density (De Young 1993). In this situation the relevant
value for $\rho_a$ is the density averaged over the volume of the radio
source.

The cocoon internal pressure is maintained by the energy carried by the
jet. For a jet power $L_j$, averaged over the radio-source lifetime
$t_{rs}$, we have $p_c \sim {2 ~t_{rs} L_j / {V_c}}$. The cocoon
volume, $V_c$, is given by $ V_c \sim 2~d_c^2~l_c$ where $d_c$ and
$l_c$ are its diameter and length.

The source age can be estimated noting that in such a cylindrical
symmetry the expansion speed decreases as $t^{-1/2}$, i.e. $v_c =
v_{c\circ}~(t / t_{rs})^{-1/2}$. Integrating the cocoon expansion from
$ t = 0$ to $ t = t_{rs}$, i.e.

$$d_c \sim 2~\int_0^{t_{rs}} ~ (t / t_{rs})^{-1/2} ~v_{c\circ} ~dt \sim
4 ~v_{c\circ} ~t_{rs} ~~,~~ t_{rs} \sim {d_c \over {4 ~v_{c\circ}}} $$

We can now express the jet power as $$L_j \sim \rho_a ~v_{c\circ}^2
~d_c^2 ~l_c ~t_{rs}^{-1} \sim 4~ \rho_a ~v_{c\circ}^3 ~d_c~l_c$$ 
Except for the external density $\rho_a$, 
the above parameters can be directly measured from our imaging and
spectroscopy data of Mrk 3. 
We have $d_c \sim 200$ pc, $l_c \sim 500$ pc and $v_{c\circ} \sim 700 $
\kms\ from which we get a jet power $$L_j \sim 2~10^{42} ~\rho_1
~~~{\rm erg~ s^{-1}}$$ where $\rho_1$ is the average density of the
unperturbed gas in units of one particle per cubic centimeter, a value
which is appropriate to the interstellar medium of our galaxy (Cox \&
Reynolds 1987).

As for the radio-source age, the value which corresponds to the
expansion law we adopted is $t_{rs} \sim 7~10^4$ years.  In reality the
cocoon lateral evolution is very complex and depends on the precise
distribution of the external medium and mass loading in the shell. The
issue of mass loading of astronomical flows has been discussed
extensively in the literature (e.g. Hartquist et al. 1986) and it will
be important in the evolution of the cocoon around the jet of Mrk 3. In
addition the bubble has now been punctured and it is in a momentum
driven phase.  A firm upper limit of $1.5 ~10^5$ years on the radio
source dynamical time scale can be set by assuming a constant expansion
at the present speed. On the other hand, a lower bound to the age of
the outflow can be set if one assumes that the ionizing photons switch
on at the same time as the radio-ejecta. The size of the Extended
Narrow Line Region of Mrk 3 ($\sim 4$ kpc, Pogge \& De Robertis 1993)
sets it to $\sim 10^4$ years. This would correspond to an expansion
speed of less than 0.1 c, similar to what is derived for powerful radio
galaxies.  

If the NLR dynamics are indeed dominated by the cocoon expansion 
the jet must have carried enough energy to accelerated the emitting gas. 
We can estimate the total kinetic energy of the high velocity gas, operatively defined
as the gas within 0\farcs2 from the radio-jet with a velocity which
differ by more than 300 \kms\ from the systemic velocity. A given
\hb\ luminosity $L_{\hb}$ corresponds to a mass of ionized gas
(Osterbrock 1989) $$M_{gas} \sim 7.5~10^{-3} ~{10^4  \over n_e}~
{L_{\hb} \over L\sun} ~~M\sun$$ Adopting an average \oiii  / \hb\ ratio
of 12.6 and an average density of 700 cm$^{-3}$
(Metz 1998), an observed flux in the \oiii\ line of $10^{-16} ~{\rm
erg~ s}^{-1}$ cm$^{-2}$ corresponds to a mass of $\sim$ 130 M\sun.
Integrating over the region of interest we find a total \oiii\ flux of
$4.3~ 10^{-13} ~{\rm erg~ s}^{-1}$ cm$^{-2}$ (about 20 \% of the total
NLR emission) and a kinetic energy of $1.2~10^{54}$ erg. Since only about
1/5 of the jet length is covered by our slits, the total kinetic
energy of the gas associated with the radio-jet is $\sim 6~10^{54}$
erg. On the other hand, during the radio-source expansion, half of the energy pumped by the jet goes into the cocoon's internal energy, while the other half
goes into accelerating the external medium. This strictly applies only
if the radiative losses of the cocoon are negligible. This is indeed
the case since the  cooling time for a temperature of $10^7$ K is $\sim
1.5 \ ~10^6\ \rho^{-1}$ years, where $\rho$ is the gas density, much
longer than the radio-source dynamical time. The total energy deposited
by the jet over the radio-source lifetime is 
$\sim 5~10^{54}$ erg, remarkably similar to the kinetic energy of the line emitting gas.
This provides {\sl quantitative} support to the idea that the gas
is accelerated (indirectly) by the jet.

\section{\label{sec:discussion} Discussion} 

Radio outflows are known to be commonly associated with Seyferts
galaxies:  in at least 50 \% of the cases the radio-emission is
extended and often shows a linear structure (Ulvestad \& Wilson 1989,
Kukula et al. 1995) with typical sizes of less than $\sim$ 1 kpc and
they rarely extend beyond 3 kpc. For Mrk 3, we estimated that the radio
source size has increased at a minimum speed of $\sim$ 3500 \kms\ ($3~10^{-3}$ pc or 8 $\mu$as per year).  Recently, HST/FOC spectroscopy has
been obtained also for NGC 1068 and NGC 4151 (Axon et al. 1998, Winge
et al. 1997). The lateral expansion of the NLR gas in these two objects
occurs at a similar speed to that in Mrk 3. In particular NGC 1068 shows a
well developed jet-cocoon structure whose transverse size is $\sim$ 100
pc. The radio-source sizes of NGC 1068 and NGC 4151 are also comparable
to that of Mrk 3 (900 and 600 pc respectively, Wilson \& Ulvestad 1987, Pedlar et al. 1993) and we therefore naturally get similar dynamical timescales to
that of Mrk 3.  This suggests that the rate of linear size expansion of
the radio-source of Mrk 3 is not unusually high, but is representative
of Seyferts with linear radio-structures. If this is the case, the
lifetime of radio outflows in Seyfert galaxies must be typically
$\lesssim 10^5$ years, otherwise we should observe Seyfert radio
sources extending well beyond 1 kpc.  As a comparison, the source age
for radio-galaxies ranges from $10^6$ and $10^8$ years (Alexander and
Leahy 1987, Liu et al 1992, Parma et al. 1998). The longer timescales
associated with radio-galaxies are also confirmed by the discovery of a
line emitting expanding shell surrounding the jet of the 3C 120  which
points an age for this source of $6 ~10^6$ years (Axon et al. 1989). It
appears that the phase of radio-activity in Seyferts is relatively
short-lived. 

Two alternative scenarios are viable at this stage. The
high nuclear luminosity and the radio-activity are causally related and
thus last for a similar time-scale in which case the whole Seyfert
phenomenon is short-lived. 
The other possibility is that the radio
activity in Seyferts is recurrent.

We can compare the radio, optical and jet luminosities of Mrk3 with
those of other classes of active galactic nuclei. Its \oiii\ luminosity
is $L_{[O III]} = 2.1~10^{42} {\rm erg~ s^{-1}}$ (Koski 1978)
while the radio-luminosity at 4.8 GHz is $L_r = 2.7~10^{30} {\rm erg
~s^{-1}~Hz^{-1}}$ (Ulvestad \& Wilson 1984; Becker, White \& Edwards,
1991) and $L_r = 2.4~10^{31} {\rm erg ~s^{-1}~Hz^{-1}}$ at 178 MHz
(Gower, Scott \& Wills 1967). The integrated radio spectral index is then 0.66 and from which we derive a radio luminosity for Mrk 3 between 10 MHz and 100 GHz of $\sim 10^{41}$ erg s$^{-1}$.

The point representing Mrk 3 in the
radio/optical luminosities plane falls within the region defined by the
{\sl radio-quiet} (RQ) quasars sample with $z < 0.5$ studied by Miller,
Rawlings \& Saunders (1991) which covers the range $10^{41} - 10^{44}
{\rm erg~ s^{-1}}$ in line luminosity.  Conversely, the radio
luminosity, $L_r$ of Mrk 3 is between 2 and 3 orders of magnitude below
the average of the {\sl radio-loud} (RL) AGN with similar line
luminosities, $L_{NLR}$ (Rawlings et al. 1989).  The Mrk 3 jet power,
$L_j$, (estimated from the expansion velocity of the emission line
region associated with the radio-jet) can be compared with estimates
based on the equipartition parameters (Rawlings \& Saunders 1991) for
RL AGNs. Mrk 3 lies again between 2 and 3 orders of magnitude below the
region defined by the objects having similar values of $L_{NLR}$.  On
the other hand the ratio between $L_j$ and $L_r$ is similar for Mrk 3
and RL AGN, indicating that there is a similar fraction of jet power
which is dissipated in radio-emission. More quantitatively,
the efficiency at which the jet power is converted in radio emission in Mrk 3 is 0.05 $\rho_1$. We speculate that the low
radio-luminosity of radio-quiet objects is due to the presence of
intrinsically less energetic outflows than in radio-loud objects,
rather than to a lower efficiency in producing synchrotron emission.
Clearly, observations of a sample of radio-quiet active nuclei with
nuclear outflows are required before firm conclusions can be drawn.

\section{Summary and conclusions}
Our HST/FOC long-slit spectroscopic observations clearly show that the
kinematics of the gas in the NLR of Mrk 3 are determined by the effects
of the interaction between the radio-outflow and the ambient gas. In
particular, along the radio-jet the line emitting gas shows two
velocity systems separated by as much as 1700 \kms\ which form broken
velocity ellipsoids.  This is a clear indication that the gas is
expanding away from the radio-jet axis and that the NLR is essentially
a cylindrical shell expanding supersonically. These results are in
close agreement with what we found in the NLR of other Seyfert galaxies
observed with HST.

The physical picture which emerges is that the high velocity shocks
induced by the outflowing plasma are heating and compressing the
external gas. Close to the radio-jets, the gas is heated to a high
temperature, $10^7$ K, and thus forms a hot cocoon surrounding the
jets. While its emission (mostly in the X-ray) can be relevant for the
ionization properties of the NLR, it is negligible for the overall
energy budget of the radio-source and thus the cocoon will undergo an
essentially adiabatic expansion.  During this expansion, it acts as a
piston, accelerating and compressing the surrounding gas which is
responsible for the narrow line emission. The jet kinetic energy is
therefore transferred to the line emitting gas after it is dissipated
in thermal energy associated with the cocoon.

With the current size of 200 pc the cocoon has expanded to several disk
scale heights. The hot gas located above the plane of the disk is
probably escaping into the halo, puncturing the bubble as indicated by
the fractures in the velocity field.  We believe that the NLR is
currently in a momentum driven phase.

Aside from the highly perturbed gas associated with the radio-jet, a
relatively quiescent component is also present, probably associated
with the galaxy disk. The radio-outflow disrupts only locally its
ordered rotation pattern. 

In the framework in which the gas motions are driven by the radio-jets,
it is possible to derive several key physical parameters describing the
properties of the radio-source associated to Mrk 3. In particular, from
the size of the cocoon and its lateral expansion speed we estimated an
upper limit to its age ($\lesssim 1.5~ 10^5$ years) and a lower limit to
the jet kinetic power ($\gtrsim 2~10^{42}$ erg s$^{-1}$).  This
provides quantitative support for the proposed NLR model, since the
total energy carried by the jet in its lifetime approximately equals
the kinetic energy associated with the perturbed high velocity NLR gas,
$\sim 6~10^{54}$ erg.  

Furthermore, if the advance speed of the Mrk 3 radio-source is taken to
be representative, it implies that radio-outflows associated with
Seyferts galaxies are short lived $\lesssim 10^5$ years, since their
typical sizes are smaller than a few kpc.

Finally, while the jet kinetic luminosity of Mrk 3 is between 2 and 3
orders of magnitude smaller than that derived for radio-loud AGNs with
similar emission line luminosity, the fraction of the jet power
dissipated in radio-emission is similar to that of RL-AGN. We thus
speculate that the main distinction between radio-quiet and radio-loud
AGN is to be ascribed to a different jet power rather than to a
different efficiency in producing synchrotron emission.

Exploring the effects of the jet propagation within the NLR of active
nuclei is clearly a very promising tool to study the jet physics and to
relate the properties of different classes of AGN. In particular it
will be important to study the dynamical timescales for a statistically
significant
number of Seyfert and radio-galaxies to further test the proposed model.

\acknowledgements

\end{document}